# Pressure induced semiconductor to metal phase transition in CsSnBr$_3$ perovskite

Md. Sajib Hossain[1], Md. Majibul Haque Babu[2,3], Tusar Saha[3], Md. Sazzad Hossain[1], Jiban Podder[3], Md. Shohel Rana[2], Abdul Barik[3], Protima Rani[3]

[1]Department of Physics, University of Dhaka, Dhaka-1000
[2]Basic Science Division, World University of Bangladesh, Dhaka-1205
[3]Department of Physics, Bangladesh University of Engineering and Technology, Dhaka-1000

## Abstract

Phase transitions in metal halide perovskites triggered by external provocations produce significantly different material properties, providing a prodigious opportunity for a comprehensive applications. In the present study, the first principles calculation has been performed with the help of density functional theory (DFT) using CASTEP code to investigate the physical properties of lead-free CsSnBr$_3$ metal halide under various hydrostatic pressures. The pressure effect is determined in the range of 0-16 GPa. Subsequently, a significant change is observed in lattice constant and volume with increasing pressure. The electronic band structure show semiconductor to metal phase transition under elevated pressure. The investigation of optical functions displays that the absorption edge of CsSnBr$_3$ perovskite is shifted remarkably toward the low energy region (red shift) with improved pressure up to 16 GPa. In addition, the absorptivity and dielectric constant also upsurges with the applied hydrostatic pressure. Finally, the mechanical properties reveal that CsSnBr$_3$ perovskite is mechanically stable and highly ductile; the ductility is increased with raising pressure. This type of semiconductor to metal phase transition may inspire a wide range of potential applications.

**Graphical diagram**

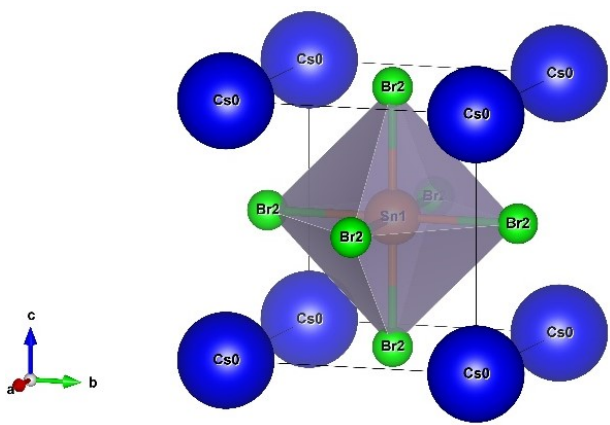

**FIG. 1.** A pristine cubic crystal structure of CsSnBr$_3$ perovskite compound

## 1. Introduction

Over the decades, the use of solar cell and optoelectronic devices has increased significantly. The material scientists are still looking for a metal that will be very highly efficient, environmentally friendly, and affordable[1]. In this concern, the scientists developed the application of perovskite-type semiconducting materials in the field of electronics such as photovoltaic cells and various optoelectronic equipment, and devices are conducted on a large scale for solar to fuel energy conversion[2-5]. Lately, the metal halide perovskites (MHPs) have reached the center of scientific interest due to some outstanding behavior such as an extensive range of absorption spectrum, intensified optical absorption, tunable bandgap, extended charge diffusion, high charge carrier mobility, and low carrier effective masses[6-7]. Besides, MHPs have adequate in the environment and economical. So, it can be easily used instead of the Si-based photovoltaic (PV) system in solar cell applications [6]. In general, MHPs have a prominent formula that is $AMX_3$, where A = a cation, M = a metal ion, and X = a halogen anion. However, lead (Pb) occupied perovskites are unfavorable and toxic for the nature [8-10]. In recent times, there has been a great deal of experimental and theoretical discussion on making effective perovskites by eliminating lead (Pb)[11-14]. Roknuzzaman et al.[11] reported that $CsSnBr_3$ metal halide perovskite reveals a considerable bandgap and high ductility as well. Consequently, it exhibits a moderate optical absorption, and is unsuitable for use as an effective solar cell. But the problem arises due to the intermediate band and bandgap transition in the metal doped halide perovskites[15,16]. The indirect bandgap can generate phonons between materials, which may create a heating effect to reduce the efficiency of optoelectronic devices[17-23].

To improve the physical properties of metal halide perovskite, researchers have been closely monitoring the effects of pressure on halide perovskites over the years since applied pressure maintains a significant effect on the physical and chemical features of halide perovskites[24-29]. Upon analyzing the reports, we have found that the lattice parameters and volumes of halide perovskite for the bulk phase is decreased with increasing pressure[26-29]. Moreover, the bandgaps of $AMX_3$ metal halide perovskites are strongly affected by the interaction of 'M' and 'X' (X=Cl, Br, I) groups, and rise with growing electronegativity of the 'X' group, which also corresponds to a diminishing M-X bond length[30]. Importantly, 'A' group does not have a strong, direct influence on the bandgap, but via the lattice parameter, mediates the M-X interactions[31]. However, there is a scarcity of investigation over metal halide perovskite by exertion of pressure. Currently, the effects of pressure on the structure and bandgap of $CsSnCl_3$, $CsGeBr_3$, $CsGeCl_3$, and $CsGeI_3$ are studied meticulously[32-33]. But the exertion of pressure on the $CsSnBr_3$ metal halide perovskite has not been studied yet.

Thus, the aim of this article, to explore the geometric structure, electronic, optical and mechanical properties of $CsSnBr_3$ extensively under different hydrostatic pressure conditions using first principles calculation by CASTEP code. Further, to find out the correlation between bond length and electronic band structure. Therefore, we believe our investigation will provide a deep insight information for future research work to explore the lead-free materials for energy harvesting applications.

2. Computational method

The structural, elastic, electronic, and optical characteristics of CsSnBr$_3$ was explored using DFT[34,35] based simulation. We conducted the computation utilizing the CASTEP (Cambridge Serial Total Energy Package) code[36,37] based plane-wave pseudopotential technique. We analyzed the band structure of CsSnBr$_3$ perovskite under the generalized gradient approximation (GGA) along with the Perdew-Burke-Ernzerhof (PBE)[38] function. For the elucidation of electron-ion interactions, we used the ultrasoft pseudopotential as a Vanderbilt type[39] . Here, we also conducted BFGS (Broyden-Fletcher-Goldfrab-Shanno) optimization technique[40]. Into the simulation, a Monkhorst–Pack[41] k point sampling of 12 × 12 × 12 for the integration of Brillouin-zone was conducted, and the plane-wave cutoff energy was settled at 550 eV. The elastic constants were also included in calculations using finite strain theory[42]. For the geometry optimization calculations, we used the convergence thresholds of 2×10−5 eV/atom for the total energy, 0.05 eV/Å for the maximum force, 0.1 GPa for maximum stress, and 0.002 Å for the maximum displacements.

3. Results and discussion
*3.1. Structural properties*

A cubic crystal structure of CsSnBr$_3$ is depicted in Fig. 1 that belongs to a cubic structure with the space group $P\,m\,\bar{3}\,m$ (#221)[43]. In addition, the unit cell contains one Cs atom, three Br atoms, and one Sn atom. In the cubic crystal, Cs atoms intrigued the corner positions with 1a Wyckof site and (0, 0, 0) fractional coordinates, Sn atoms take place at the body-centered position with 1b Wyckof site and (0.5, 0.5, 0.5) fractional coordinates, and the Br atoms is at the face-centered positions with 3c Wyckof site[11] and (0, 0.5, 0.5) fractional coordinates. The calculated lattice parameter is faintly higher than the experimental data. Table I shows the evaluated equilibrium lattice constant and the corresponding analyzed unit cell volume in this simulation with theoretical and experimental outcomes of the cubic CsSnBr$_3$. Importantly, we have done the simulation under various hydrostatic pressures from 0 to 16 GPa with a step of 2 GPa. The computed lattice parameter at 0 GPa in this simulation fits nicely with the previous theoretical value[62]. It is observed that Cs-Cs has significantly larger bond length than Cs-Cl and Cl-Sn in PBE method. The value of lattice parameter, cell volume and bond lengths (Cs-Br, Cs-Cs and Br-Sn) reduces in a readily way with the pressure's growth; consequently, the space between atoms tends to shrink. For this reason, the repulsive force between atoms becomes better which conducts to the solidity of crystal compression under raised pressure. The impact of given pressure on the lattice constant, unit cell volume and bond lengths are given in Fig. 2 (a-c).

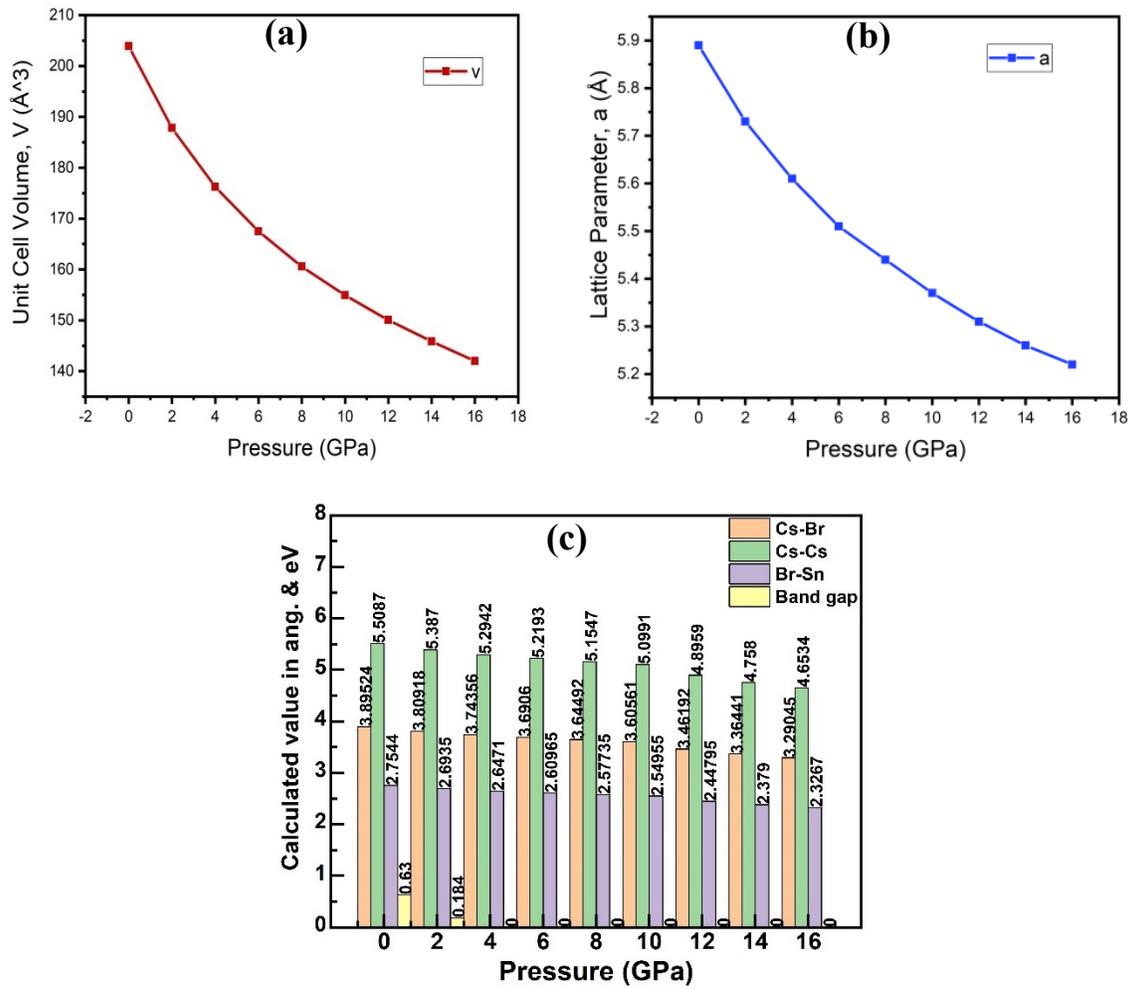

**FIG. 2.** An enumerated (a) unit cell volume (V), (b) lattice parameter ($a$) and (c) bond length stimulation of CsSnBr$_3$ metal halide under enhanced pressure.

**TABLE I.** The simulated data of lattice parameter ($a$), and unit cell volume of CsSnBr$_3$ at variant pressures

| Pressure | Lattice constant in (Å) | Volume in Å$^3$ |
|---|---|---|
| 0 | 5.89 | 203.92 |
| 2 | 5.73 | 187.83 |
| 4 | 5.61 | 176.25 |
| 6 | 5.51 | 167.51 |
| 8 | 5.44 | 160.58 |
| 10 | 5.37 | 154.92 |
| 12 | 5.31 | 150.05 |
| 14 | 5.26 | 145.86 |
| 16 | 5.22 | 141.9 |

*3.2. Electronic properties*

We have explored electronic properties such as band structure and total density of state (TDOS) for understanding the phase transition (semiconductor to metal) of CsSnBr$_3$ perovskite. The simulated electronic band structure and TDOS under various hydrostatic pressures utilizing generalized gradient approximation (GGA) of Perdew-Berke-Ernzerhof (PBE) are displayed in Fig. 3. According to the semi-conductive theory, the band near the Fermi level is noteworthy for knowing the material's physical behavior. Without introducing any external pressure at R-point, the bandgap of CsSnBr$_3$ is 0.630 eV. The calculated direct bandgap of the GGA approach is far away from the experimental bandgap value (1.75 eV)[44] and it happens due to the GGA approach. It is concluded that GW method[45], hybrid method[46] can improve the bandgaps in these systems. Comprehensive deportment of the discrepancy in the bandgap ($E_g$) and the band structure with applied pressure is distinct from functional employed and that the PBE proposition issued reasonably precise results, which proposed the operation of GGA-PBE functional for pressure inquiry on materials[47,48].

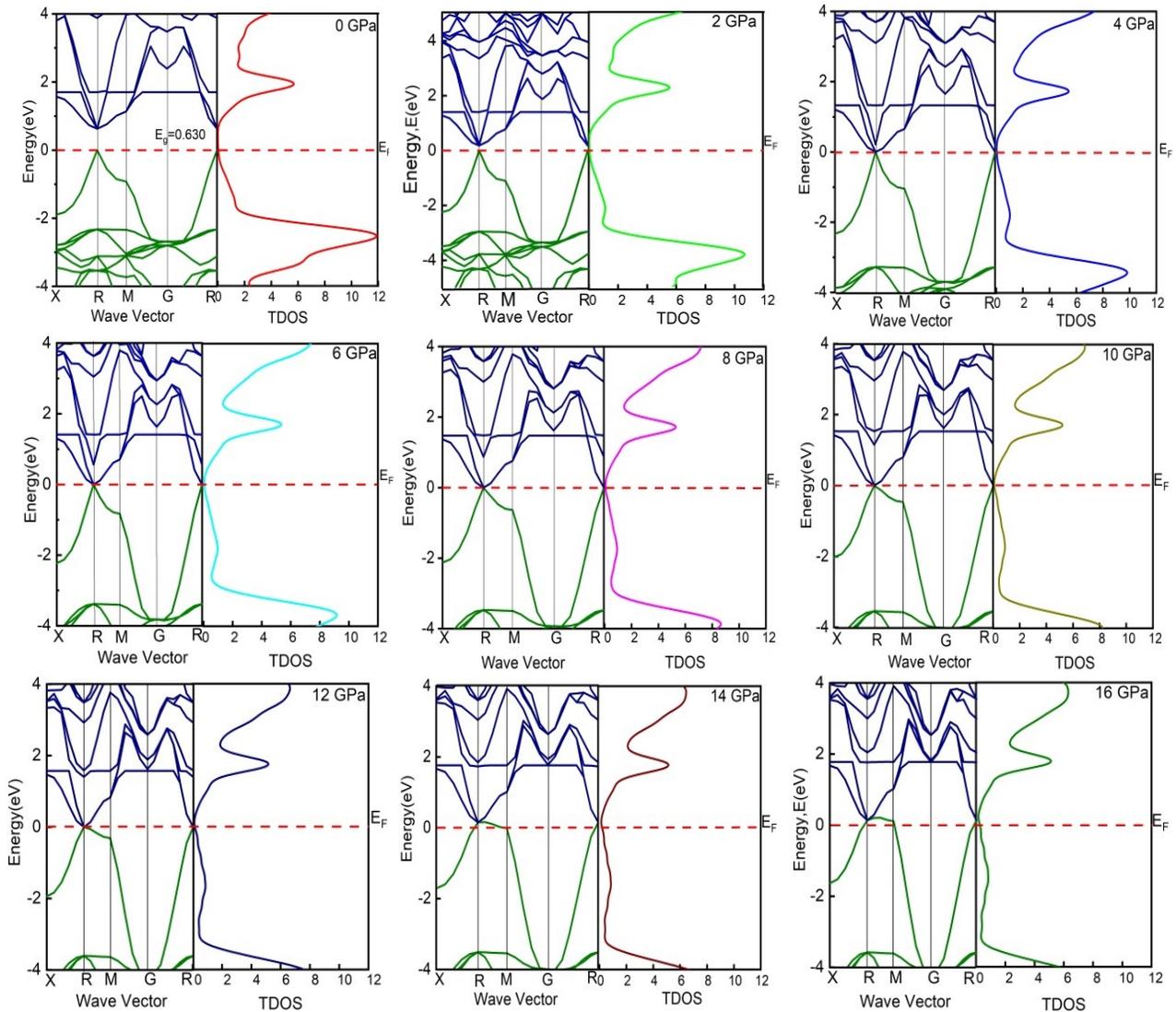

**FIG. 3.** Calculated bandgap with TDOS of CsSnBr$_3$ under different pressures.

For this reason, we have calculated the band structure of CsSnBr$_3$ perovskite under GGA along with PBE. We have calculated the band structure with a step of 2 GPa up to 16 GPa. With increasing pressure, the valance band maxima (VB) and conduction band minima (CB) at R-point start to shift toward E$_F$. In Fig. 3, there is no bandgap for 4GPa, 6GPa, 8GPa, 10GPa, and 12 Gpa, respectively. It is also seen (Fig. 3) that valence band maximum (VBM) and conduction band minimum (CBM) is reached at the Fermi level for 14 GPa which means semiconductor to metallic transition[49]. Hence, a metallic behavior of CsSnBr$_3$ has emerged. By analyzing TDOS, we can understand the metallic behavior of CsSnBr$_3$ correctly. In this way, Fig. 3 shows that the value of TDOS is zero at the Fermi level between 0 and 2 GPa. Afterward, we notice a negligible value at Fermi level above 2 GPa hydrostatic pressures. Interestingly, we find non-zero TDOS Fermi level ranges from 14 to 16 GPa pressures. Note that a finite value of TDOS of the material at a specific pressure indicates that material may undergo a metallic transition on that pressure[49,50]. Therefore, a semiconducting to metallic transition of CsSnBr$_3$ has firmly confirmed by 14 to 16 GPa hydrostatic pressure.

We also find that the CsSnBr$_3$ perovskite becomes metal when reach at a certain bond length. It is found metal when pressure is introduced in the range from 0 to 16 GPa, then the bond lengths of Cs-Br, Cs-Cs and Br-Sn is obtained from 4.16203 to 3.72221, 5.886 to 5.264, and 2.943 to 2.632 Å respectively. Therefore, we can say that the semiconducting nature of CsSnBr$_3$ is shifted to metallic nature when bond lengths of Cs-Br, Cs-Cs and Br-Sn reach under 3.72221, 5.264 and 2.632 Å. Throughout this study, we have meticulously analysed that a metallic behaviour and certain bond length of CsSnBr$_3$ is emerged by 14 GPa hydrostatic pressure. For getting the semiconducting-metallic transition point of CsSnBr$_3$, future experimental exploration should be carried out. We firmly expect that our analysis will be helpful for future experimental exploration.

*3.3. Optical properties*

It can reliably comprehend the optical behavior of a material by observing the electronic configuration. Generally, optical properties measured by fascinating behavior of photon energy. The optical properties of CsSnBr$_3$ in addition to the real and imaginary part of dielectric functions, absorption, conductivity, and reflectivity, have explored in different hydrostatic pressure (from 0Gpa to 16 GPa). The CsSnBr$_3$ is not good for solar application due to its low optical absorption and conductivity according to Roknuzzaman, M. et al[11] literature. By applying different hydrostatic pressure, we can promote the execution of CsSnBr$_3$ as solar cell and other optoelectronic tools implementation.

The absorption coefficient ($\alpha$) is the most important parameter that measures the ability of a material to absorb incident photons and the capability to achieve optimum solar energy conversion. This is the effective parameter for evaluating the effectiveness of a material in photovoltaic devices (e.g., solar cells). In generally, the intra-band transitions can produce the low energy infrared spectra, whereas the inter-band contributions provide the high energy absorption spectra. We have shown Fig. 4 the optical absorption of CsSnBr$_3$ under various hydrostatic pressures. We have analyzed the potentiality of a material to absorb light energy from the optical absorption coefficient, which gives important statistics about the material's solar energy reformation efficiency. Using the optical absorption coefficient of a material, we can explain the penetration of light at specific energy(wavelength) into the material before absorbed[51]. We have exhibited the absorption spectra of CsSnBr$_3$ as a function of photon energy under variant hydrostatic pressure up to 16 GPa. Fig. 4 (a) shows that the absorption edge is moved towards the shorter energy region (redshift) with increasing in pressure. Its absorption increased rapidly by applying various hydrostatic pressures and towards a prominent range in the visible and ultraviolet region. From Fig. 4, we see that the maximum broad absorption pick lies in the ultraviolet zone that the CsSnBr$_3$ metal halide would be a fruitful material to decontaminate surgical apparatuses. By analyzing the light energy absorption in the material's largest ultraviolet region, the decontaminated surgical instrument made with it can be accurately determined[52].

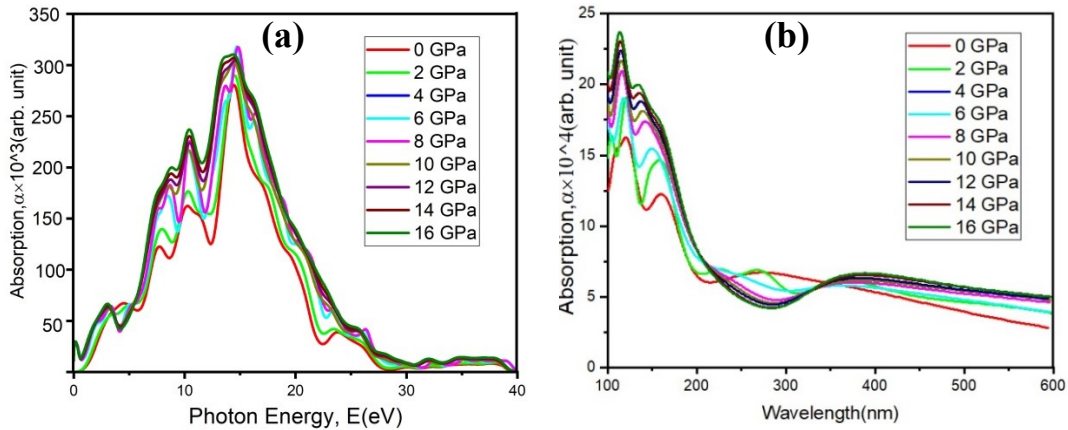

**FIG. 4.** Computed absorption profile of CsSnBr$_3$ perovskite under variant pressures (a) as a function of photon energy (eV) and (b) as a function of wavelength

We have computed the wavelength-dependent absorption coefficient to better estimate the light absorbance complexion of CsSnBr$_3$ in the visible area under increasing pressure. Without applying hydrostatic pressure, very small absorption has shown in Fig. 4 (b). Nevertheless, the absorption coefficient enhances with increasing pressure. Under pressure, CsSnBr$_3$ can be a satisfactory substitution for contaminated Pb-including materials.

The reflectivity of a material describes the potentiality of a material to reflect the coming photon energy of the surface of the CsSnBr$_3$. Under different hydrostatic pressure, we have shown the reflectivity spectra of the CsSnBr$_3$ for photon energy till 30 eV given in Fig. 5. Reflectivity of CsSnBr$_3$ has increased with increasing in various hydrostatics pressures, which reduces the strength of the solar cell. A further study is required to reduce the reflectivity of the pressure convinced CsSnBr$_3$ in the visible energy region, which may intensify the absorptivity and solar cell productivity.

The investigated optical conductivity (real part) is exhibited in Fig. 5, which is accountable for photoconductivity[53]. Upon applying various hydrostatic pressures, the halide compound $CsSnBr_3$ increased their conductivity due to enhanced photon absorption. This result supported the prediction from Band Structure and Density of States calculations. Analyzing the dielectric function is essential to exploring the charge-recombination rate and the proficiency of optoelectronic instruments[54]. The photoconductivity (electrical conductivity) of halide perovskite $CsSnBr_3$ increased with increasing photon energy by applying various hydrostatic pressures.

The behavior of a material under incident light is defined as a dielectric function, and the value of the dielectric function at zero photon energy is referred to as the static dielectric function. We have calculated the real and imaginary parts of the dielectric function (Fig. 5) up to photon energy of 20eV. The stable peak summit of the dielectric constant of both real and imaginary sides of the $CsSnBr_3$ perovskite develops in the visible region with intensified pressure, as demonstrated in Fig. 5. The materials with wide-ranging bandgap show a less stable value of dielectric constant[55]. In addition, the real ($\varepsilon 1$) and imaginary parts ($\varepsilon 2$) of the dielectric function for the pressure induced compound $CsSnBr_3$ exhibited a sharp peak at the low energy region. The imaginary part is entirely related to the bandgap and total density of state and can explain the absorption nicely[56].

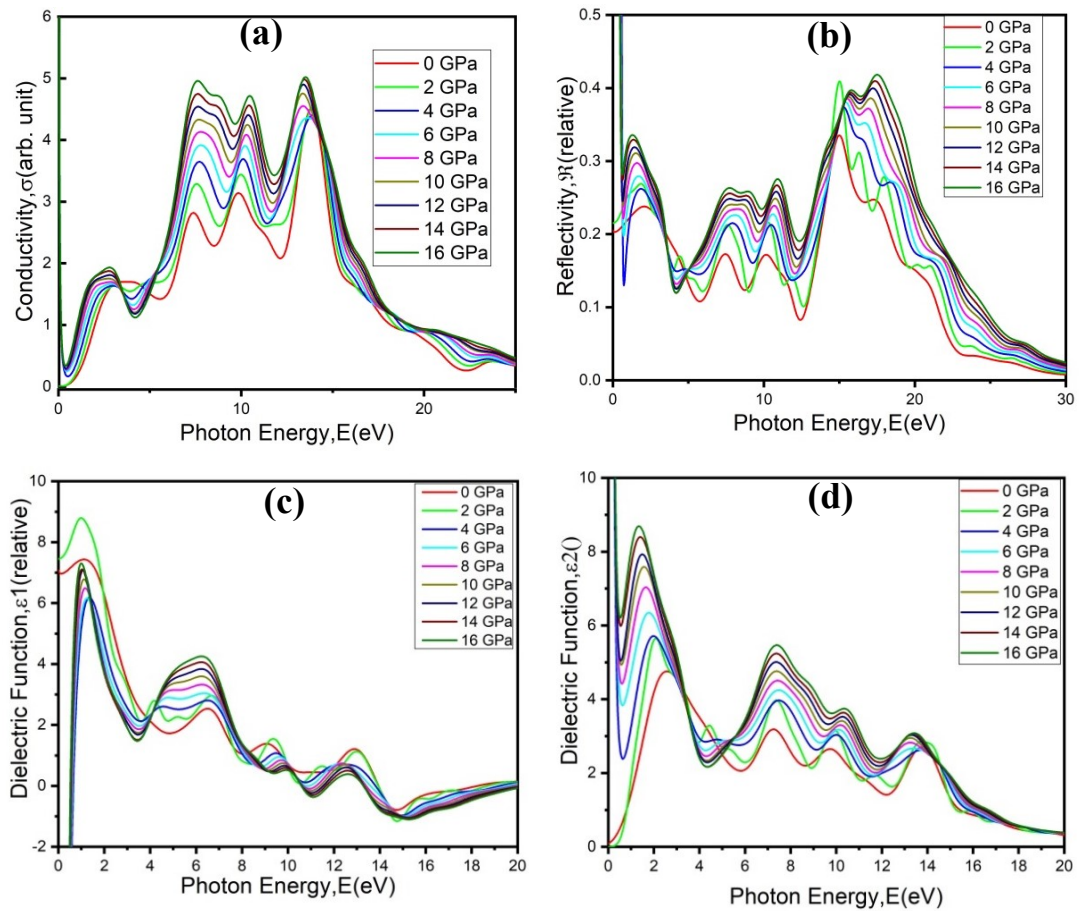

**FIG. 5.** Simulated pressure applied bands of (a) optical conductivity (b) reflectivity (c) real part of dielectric function, and (d) imaginary part of the dielectric of CsSnBr$_3$ perovskite

At a high energy segment (above 26 eV) for all the pressured-enhanced CsSnBr$_3$ samples, the imaginary region of dielectric constant moves toward zero, and the real segment goes to unity. This outcome obeys that all the pressure-increasing data divulge better transparency and consequently minimum absorption in the high energy segment (above 26 eV), which is also axiomatic from the absorption coefficient profile as depicted in Fig. 3 (a). The investigation of CsSnBr$_3$ under different pressures shows high transparency and low absorption in the long-energy segment (above 26 eV) due to the minimum dielectric outcome.

### 3.4. Mechanical properties

Mechanical properties typically are essential to know the mechanical steadiness of the material properly. So, we have studied the mechanical properties throughout finite strain theory[57]. Analyzing elastic constants, we can know dynamic statistics about the potentiality of a crystal to combat outside pressure. The CsSnBr$_3$ perovskite has three elastic constants, which are $C_{11}$, $C_{12}$, and $C_{44}$. The computed elastic constants in different pressures are given in Table II. Using Born stability criteria[58] $(C_{11} + 2C_{12}) > 0$, $C_{44} > 0$, and $(C_{11} - C_{44}) > 0$, we can understand the steadiness of a crystal. From Table II, we find that the CsSnBr$_3$ assures the Born steadiness basis. Therefore, the metal halide is mechanically stable under different pressure. In addition, it is also seen (Table II) that the data of $C_{11}$ and $C_{12}$ enlarge speedily with enhancing pressures. However, the value of $C_{44}$ remains almost the same under increasing pressure, up to 16 GPa. From the Cauchy pressure ($C_{11}$-$C_{44}$), the ductile and brittle nature of materials is estimated. Negative data of Cauchy reveals that it is brittle in characteristics.

**TABLE II.** The simulated data of $C_{ij}$ (GPa) and Cauchy pressure $C_{12}$–$C_{44}$ (GPa) of CsSnBr$_3$ metal halide under form pressure.

| Pressure (GPa) | $C_{11}$ | $C_{12}$ | $C_{44}$ | $C_{12}$-$C_{44}$ |
|---|---|---|---|---|
| 0 | 43.51 | 7.36 | 5.16 | 2.2 |
| 2 | 61.88 | 10.38 | 5.30 | 5.08 |
| 4 | 80.09 | 13.97 | 5.39 | 8.53 |
| 6 | 97.23 | 17.26 | 5.39 | 11.87 |
| 8 | 113.25 | 20.57 | 5.40 | 15.17 |
| 10 | 128.33 | 23.75 | 5.37 | 18.38 |
| 12 | 143.11 | 27.08 | 5.33 | 21.75 |
| 14 | 157.46 | 30.57 | 5.25 | 25.32 |
| 16 | 171.29 | 34.05 | 5.13 | 28.92 |

Here, the Cauchy pressure of the CsSnBr$_3$ metal halide under all calculated pressure is greater than zero and enhances with increasing pressure. So, CsSnBr$_3$ perovskite discloses a ductile mechanical behavior. However, the mechanical properties including Bulk modulus (B=($C_{11}$ +

$2C_{12})/3$), Shear modulus (G=$\frac{1}{5}(3C_{44} + C_{11} - C_{12})$), Young's modulus (E=$\frac{9B_0G}{3B_0+G}$), Pugh's ratio(B/G), and Poisson's ratio($v=-1+\frac{Y}{2G}$), of the CsSnBr$_3$ metal halide, are explored using the Voigt–Reuss–Hill (VRH) averaging method[59]. The calculated values are given in Table III.

The Pugh's ratio is an essential sector to understand the ductile and brittle nature of a metal halide. The minimum (high) value of B/G says the brittle (ductile) behavior of the material, and the critical value is known as 1.75[60]. Table III shows that at zero pressure, the pugh's ratio (B/G) of CsSnBr$_3$ is larger than the critical value, indicating the perovskite's ductile behavior. It is also showed that the pugh's ratio (B/G) improves with the enhanced pressure, which says that the improvement of applied pressure can develop the ductility of CsSnBr$_3$. By analyzing Poisson's ratio (v), we can understand a cubic crystal's bonding forces and steadiness. The highest value of $v$ is 0.5, and the lowest value is 0.25 of $v$, which can help us study the central forces in the cubic crystals[61]. The interatomic forces of charged crystals are known as central forces[62]. Table III shows that the value of Poisson's ratio (v) of CsSnBr$_3$ perovskite without pressure is 0.274, which between 0.5 and 0.25, unveiling the alive of central forces in the perovskite. The simulated value of $v$ improves with expanding pressure. Nevertheless, $V$ does not change much after 14 GPa that reveals powerful central forces subsist in CsSnBr$_3$. Throughout Poisson's ratio, we can understand the brittleness and ductility of the CsSnBr$_3$ perovskite soundly. We get that the critical value of $v$ is 0.2614, which is the indicator of the ductile and brittle nature of a material. Investigating table 3, we see that at zero GPa pressure, Poisson's ratio of CsSnBr$_3$ metal is larger than the critical value, which indicates the ductile behavior of the organic metal halide. So, we can enhance the ductility by introducing hydrostatic pressure. The discrepancy of B/G and v of CsSnBr$_3$ metal halide with variant pressure have been depicted in Fig. 6 (a-b). It is manifested that the ductility of CsSnBr$_3$ develops with increasing pressures. So, the hydrostatic pressure is a very significant approach for the fabrication of high ductility CsSnBr$_3$ devices.

**TABLE III.** The simulated mechanical properties of CsSnBr$_3$ perovskite

| Pressure (GPa) | B (GPa) | G (GPa) | Y (GPa) | B/G | Poisson's ratio, v |
|---|---|---|---|---|---|
| 0 | 19.41 | 10.33 | 26.32 | 1.88 | 0.274 |
| 2 | 27.55 | 13.48 | 34.77 | 2.04 | 0.290 |
| 4 | 36.01 | 16.46 | 42.85 | 2.19 | 0.302 |
| 6 | 43.84 | 19.23 | 50.33 | 2.28 | 0.309 |
| 8 | 51.46 | 21.78 | 57.26 | 2.36 | 0.315 |
| 10 | 58.61 | 24.14 | 63.68 | 2.43 | 0.319 |
| 12 | 65.76 | 26.40 | 70.17 | 2.49 | 0.322 |
| 14 | 72.87 | 28.53 | 75.71 | 2.55 | 0.327 |
| 16 | 79.80 | 30.53 | 81.23 | 2.61 | 0.330 |

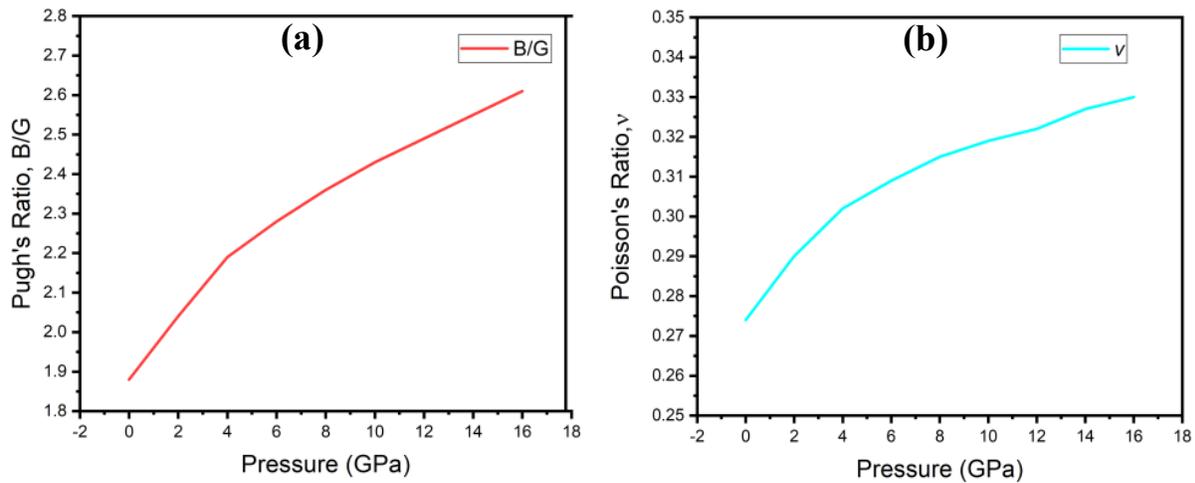

**FIG. 6.** (a) Discrepancy of Pugh's ratio (b) discrepancy of Poisson's ratio of $CsSnBr_3$ metal halide at variant pressure

## 4. Conclusion

Throughout exploring the formation, optical, electronic, and elastic properties under variant hydrostatic pressure using DFT-based CASTEP code, we get essential information on cubic $CsSnBr_3$ perovskite. We see that the perovskite's lattice parameter and unit cell volume reduce with pressure, but the elastic moduli increase with increasing pressure. That obeys the hardness of $CsSnBr_3$. According to the analysis of Poisson's ratio and Pugh's ratio, the $CsSnBr_3$ material exhibits a developing affinity of ductility with raising pressure. So, for high ductility devices application, we can use the material. By investigating band structure under variant pressure, we see the semiconducting to the metallic transition of the perovskite. Further exploration of the optical absorption and conductivity under elevated pressure suggests that the cubic $CsSnBr_3$ metal halide can be very useful as a photovoltaic cell and various optoelectronic devices.

**Author's credit**

**Md. Sajib Hossain:** Simulate, Investigation, writing original draft preparation; **Md. Majibul Haque Babu:** Visualization, Investigation, writing original draft preparation, Reviewing and Editing; **Jiban Podder:** Supervision, Conceptualization, Reviewing and Editing; **Md. Sazzad Hossain:** Formal analysis, Investigation, Reviewing and Editing; **Md. Shohel Rana:** Methodology, Formal analysis; **Abdul Barik:** Validation, Investigation; **Tusar Saha**: Formal analysis (drawing the graph); **Protima Rani**: Validation, Investigation